\documentstyle[12pt]{article}

\newcommand{\mtx}[2]{\left(\begin{array}{#1}#2\end{array}\right)}

\begin{document}

\begin{center}

\bigskip
{\Large Distributed Entanglement}\\

\bigskip

Valerie Coffman,$^{(1)}$ Joydip Kundu,$^{(2)}$

and William K.~Wootters$^{(3)}$\\

\bigskip

{\small{\sl

$^{(1)}$Department of Physics, Johns Hopkins University, 
Baltimore, MD 21218, USA \vspace{2mm}

$^{(2)}$Department of Physics, MIT, Cambridge, MA 02139, 
USA \vspace{2mm}

$^{(3)}$Department of Physics, Williams College, Williamstown, 
MA 01267, USA and Isaac Newton Institute, 
University of Cambridge, Cambridge, CB3 0EH, UK}}\vspace{3cm}

\end{center}
\subsection*{\centering Abstract}
{Consider three qubits A, B, and C which may be entangled with 
each other.  We show that there is 
a trade-off between A's entanglement with B and its entanglement
with C.  This relation is expressed 
in terms of a measure
of entanglement called the ``tangle,'' which is related to
the entanglement of formation.  Specifically, we show that
 the tangle
between A and B, plus the tangle between A and C, cannot be
greater than the tangle between A and the pair BC.  This 
inequality is as strong as it could be, in the sense that
for any values of the tangles satisfying the corresponding
equality, one can find a quantum state consistent with those
values.  Further exploration of this result leads to a 
definition of the ``three-way tangle'' of the system,
which is invariant under permutations of the qubits.  }

\vfill

PACS numbers: 03.65.Bz, 89.70.+c\vfill

\newpage

Quantum entanglement has rightly been the subject of much
study in recent years as a potential resource
for communication and information processing.  
As with other resources such
as free energy and information,
one would like to have a quantitative theory of
entanglement giving specific rules about how it can
and cannot be manipulated; indeed such a theory 
has begun to be developed.
The first step in building the theory has been to quantify
entanglement itself.  In the last few years a number of 
entanglement measures
for bipartite states have been introduced 
and analyzed [1--7],
the one most relevant to the present work
being the ``entanglement of formation'' \cite{entmeasures}, 
which is intended to
quantify the amount of quantum communication required to create a
given state.  In the present paper we draw on some of the earlier 
work on entanglement
of formation \cite{Hill,Wootters} in order to explore another basic 
quantitative question: 
To what extent does entanglement between two objects limit
their entanglements with other objects?

Unlike classical correlations, quantum entanglement cannot be
freely shared among many objects.  For example, 
given a triple of spin-$\frac{1}{2}$ particles
A, B and C, if particle A is fully
entangled with particle B, {\em e.g.}, 
if they are in the singlet state
$\frac{1}{\sqrt{2}}(|\hspace{-0.7mm}\uparrow\downarrow\rangle 
- |\hspace{-0.7mm}\downarrow\uparrow\rangle )$,
then particle A cannot be simultaneously entangled 
with particle C.  (If A were
entangled with C, then the pair AB would also be entangled 
with C and would therefore have a mixed-state density matrix,
whereas the singlet state is pure.)  One expects that a less
extreme form of this restriction should also hold: if A is
{\em partly} entangled with B, then A can have only a limited
entanglement with C.  The first goal 
of this paper is to verify this intuition and express it
quantitatively.  We will see that the restriction
on the sharing of entanglement takes a particularly elegant
form in terms of a measure of entanglement called the ``tangle,''
which is closely related to the entanglement of formation.
Further analysis of this result will lead us naturally to
a quantity that measures a three-way entanglement
of the system and is invariant under all permutations
of the particles \cite{multi}.

The present work is related to recent work 
on the characterization of multiparticle
states in terms of invariants under local 
transformations [9--12]; indeed, both the
tangle and our measure of three-way entanglement are
invariants in this sense.  Our work is also related to research 
exploring the connection between
entanglement and cloning [13--17].
An example along these lines was studied by 
Bru\ss, who asked, in the case of a singlet
pair AB, to what extent particle B's
entanglement with particle A can be shared symmetrically 
and isotropically
with a third particle, for a purpose such as teleportation
where isotropy is desired \cite{Bruss}.  
Our investigation is similar in spirit to that of Bru\ss \, 
but has a different focus in that we are 
looking for a general law governing the 
splitting or sharing
of entanglement; thus, for example, we make no assumptions 
about the symmetry of the system.
Some of the results presented here have been mentioned in
a recent paper by one of us \cite{Royal}, but the proofs 
and most of the 
details and observations have not been previously published.
In this paper we confine our attention to binary 
quantum objects (qubits) such as spin-$\frac{1}{2}$ 
particles---we will use the generic basis labels $|0\rangle$
and $|1\rangle$ rather than $|\hspace{-0.1cm}\uparrow\rangle$ and 
$|\hspace{-0.1cm}\downarrow\rangle$---but the same questions 
could be raised
for larger objects.   

We begin by defining the tangle.  Let A and B be a pair of 
qubits, and let the density matrix of the pair be $\rho_{AB}$, which
may be pure or mixed.  We define the ``spin-flipped'' density
matrix to be 
\begin{equation}
\tilde{\rho}_{AB} = 
(\sigma_y \otimes \sigma_y) \rho^*_{AB} (\sigma_y \otimes \sigma_y),
\end{equation}
where the asterisk denotes complex conjugation in the standard 
basis
$\{|00\rangle$, $|01\rangle$, $|10\rangle, |11\rangle\}$
and $\sigma_y$ expressed in the same basis is the matrix
$\mtx{cc}{0&{-i}\\i&0}$.
As both $\rho_{AB}$ and $\tilde{\rho}_{AB}$ are positive operators,
it follows that the product $\rho_{AB} \tilde{\rho}_{AB}$, though
non-Hermitian, also has only real and non-negative eigenvalues.  
Let the
square roots of these eigenvalues,
in decreasing order, be $\lambda_1$, $\lambda_2$, $\lambda_3$,
and $\lambda_4$.  Then the tangle of the density matrix $\rho_{AB}$
is defined as 
\begin{equation}
\tau_{AB} = 
[{\rm max}\{\lambda_1 - \lambda_2 -\lambda_3 -\lambda_4,0\}]^2.
\label{tau}
\end{equation}
For the special case in which the state of AB is pure, the matrix
$\rho_{AB} \tilde{\rho}_{AB}$ has only one non-zero eigenvalue, and
one can show that $\tau_{AB}=4\det \rho_{A}$, where $\rho_{A}$
is the density matrix of qubit A, that is, the trace of $\rho_{AB}$
over qubit B.

It is by no means obvious from the definition that the tangle
is a measure of entanglement for mixed states.  
This interpretation comes from
previous work, in which a specific connection is established
between the tangle and
the entanglement of formation of a pair of qubits \cite{Wootters}.
For the purpose of this paper it is sufficient to note that
$\tau = 0$ corresponds to an unentangled state, $\tau = 1$ 
corresponds to a completely entangled state, and 
the entanglement of formation is a 
monotonically increasing function of $\tau$.\footnote{The 
entanglement of formation is given by 
$E=h(\frac{1}{2} + \frac{1}{2}\sqrt{1-\tau})$, where
$h$ is the binary entropy function 
$h(x)= -x\log x -(1-x)\log (1-x)$.} 
(The earlier work did not define the tangle 
{\em per se} but rather the ``concurrence,''
which is simply the square root of the tangle.  It is with some 
hesitation that we introduce the new term ``tangle'' here instead
of speaking of the square of the concurrence.  However, 
for our present purpose the tangle does seem to be the more natural
measure to use, and the discussion is significantly
simplified by introducing this term.)
At present, 
the tangle is defined only for a pair of qubits,
not for higher-dimensional systems.  

We now turn to the first problem of this paper: given a pure
state of three qubits A, B, and C, how is the tangle between
A and B related to the tangle between A and C?
For this special case---a pure state of three
qubits---the formula for the tangle
simplifies: each pair of qubits, being
entangled with only one other qubit in a joint pure state,
is described by a density matrix having 
at most two non-zero eigenvalues.  It follows that the
product $\rho_{AB} \tilde{\rho}_{AB}$ also has only
two non-zero eigenvalues.  We can use this fact and
Eq.~(\ref{tau}) to write the 
following inequality for the tangle $\tau_{AB}$ between 
A and B.
\begin{eqnarray}
\tau_{AB} & = & (\lambda_1 - \lambda_2)^2
 = {\lambda_1}^2 + {\lambda_2}^2 - 2\lambda_1 \lambda_2  
 \nonumber\\
 & = & {\rm Tr} (\rho_{AB}\tilde{\rho}_{AB}) - 
2\lambda_1 \lambda_2
          \le {\rm Tr} (\rho_{AB}\tilde{\rho}_{AB}). 
  \label{2lambdas}
\end{eqnarray}
Here $\rho_{AB}$ is the density matrix of the pair AB, obtained from
the original pure state by tracing over qubit C.  
Eq.~(\ref{2lambdas})
and the analogous equation for $\tau_{AC}$ allow us to bound the sum
$\tau_{AB} + \tau_{AC}$:
\begin{equation}
\tau_{AB} + \tau_{AC} \le 
{\rm Tr} (\rho_{AB}\tilde{\rho}_{AB}) + 
{\rm Tr} (\rho_{AC}\tilde{\rho}_{AC}).   \label{tautau}
\end{equation}
The next paragraph is devoted to evaluating  
the right-hand side of this inequality.

Let us express the pure state $|\xi\rangle$
of the three-qubit system
in the standard basis $\{|ijk\rangle \}$, where each
index takes the values 0 and 1:                         
\begin{equation}
|\xi\rangle = \sum_{ijk} a_{ijk}|ijk \rangle.
\end{equation}
In terms of the coefficients $a_{ijk}$, we can write
${\rm Tr} (\rho_{AB}\tilde{\rho}_{AB})$ as                    
\begin{equation}
{\rm Tr} (\rho_{AB}\tilde{\rho}_{AB})
=\sum a_{ijk}a^*_{mnk}\epsilon_{mm^\prime}\epsilon_{nn^\prime}
a^*_{m^\prime n^\prime p}a_{i^\prime j^\prime p}
\epsilon_{i^\prime i}\epsilon_{j^\prime j},
\end{equation}
where $\epsilon_{01}=-\epsilon_{10}=1$ and   
$\epsilon_{00}=\epsilon_{11}=0$ and the sum is over all the
indices.  We now replace the product 
$\epsilon_{nn^\prime}\epsilon_{j^\prime j}$ with the 
equivalent expression
$\delta_{nj^\prime}\delta_{n^\prime j}-
\delta_{nj}\delta_{n^\prime j^\prime}$, and in the first of
the two resulting terms (that is, the one associated with
$\delta_{nj^\prime}\delta_{n^\prime j}$)
we perform a similar substitution for
$\epsilon_{mm^\prime}\epsilon_{i^\prime i}$.  These substitutions
directly give us 
\begin{equation}
{\rm Tr} (\rho_{AB}\tilde{\rho}_{AB}) = 2\det {\rho_A}
-{\rm Tr} (\rho_B^2) + {\rm Tr} (\rho_C^2), \label{dettr}
\end{equation}
where $\rho_A$, $\rho_B$, and $\rho_C$ are the 2x2 density
matrices of the individual qubits.  Because each of these
matrices has unit trace, we can rewrite Eq.~(\ref{dettr})
as
\begin{equation}
{\rm Tr} (\rho_{AB}\tilde{\rho}_{AB}) = 
2(\det{\rho_A} + \det {\rho_B} - \det {\rho_C}).
\end{equation}
By symmetry we must also have 
\begin{equation}
{\rm Tr} (\rho_{AC}\tilde{\rho}_{AC}) = 
2(\det {\rho_A} + \det {\rho_C} - \det {\rho_B}).
\end{equation}
Summing these last two equations, we finally get a simple expression
for the right-hand side of Eq.~(\ref{tautau}), namely,
\begin{equation}
{\rm Tr} (\rho_{AB}\tilde{\rho}_{AB})
+ {\rm Tr} (\rho_{AC}\tilde{\rho}_{AC}) = 4\det {\rho_A}. \label{trdet}
\end{equation}
Eqs.~(\ref{tautau}) and (\ref{trdet}) give us our first main result:
\begin{equation}
\tau_{AB}+\tau_{AC} \le 4\det\rho_A. \label{first}
\end{equation}

We can interpret the right-hand side of Eq.~(\ref{first}) as follows.  
Thinking of the pair BC as a single object, it makes sense to
speak of the tangle between qubit A and the pair BC,
because, even though 
the state space of BC is 
four-dimensional, only two of those dimensions 
are necessary to
express the pure state $|\xi\rangle$ of ABC.  (The two necessary
dimensions are those spanned by the two eigenstates of $\rho_{BC}$ 
that have non-zero eigenvalues.  That there are only two such eigenvalues
follows from the fact that A is only a qubit.)
We may thus treat A and BC, at least for this purpose, 
as a pair of qubits, and
so the tangle is well defined.  In fact this tangle 
$\tau_{A(BC)}$ 
is simply $4\det{\rho_A}$ as we have mentioned before.  We
can therefore rewrite our result as
\begin{equation}
\tau_{AB}+\tau_{AC} \le \tau_{A(BC)}.  \label{3taus}
\end{equation} 
Informally, Eq.~(\ref{3taus}) can be expressed as follows.  Qubit 
A has a certain amount of entanglement with the pair BC.
This amount bounds A's entanglement with qubits B and C
taken individually, and the part of the entanglement 
that is devoted to qubit B (as measured by the tangle) 
is not available to qubit C. 

We will say more shortly about the case of three qubits in a pure
state, but at this point it is worth mentioning a generalization
to mixed states.  If ABC is in a mixed state $\rho$, then 
$\tau_{A(BC)}$ is not defined, because all four dimensions of
BC might be involved, but we can define a related
quantity $\tau^{min}_{A(BC)}$ via the following prescription.  
Consider all possible
pure-state decompositions of the state $\rho$, that is, all
sets $\{(\psi_i,p_i)\}$ such that $\rho = \sum_i p_i 
|\psi_i\rangle\langle\psi_i|$.
For each of these decompositions, one can compute the average
value of $\tau_{A(BC)}$:
\begin{equation}
\langle \tau_{A(BC)} \rangle = \sum_i p_i \tau_{A(BC)}(\psi_i).
\end{equation}
The minimum of this average over all decompositions of $\rho$ is 
what we take to be $\tau^{min}_{A(BC)}(\rho)$.
The following analogue of Eq.~(\ref{3taus}) then holds for mixed 
states:
\begin{equation}
\tau_{AB}+\tau_{AC} \le \tau^{min}_{A(BC)}.  \label{mixed}
\end{equation}
To prove this, consider the pure 
states $|\psi_i\rangle$ belonging to an 
optimal decomposition of $\rho$, that is, a decomposition that 
minimizes
$\langle\tau_{A(BC)}\rangle$.  We can write our basic inequality,
Eq.~(\ref{3taus}), for each such pure state and then
average both sides
of the inequality over the whole decomposition.  The right-hand
side of the resulting inequality is $\tau^{min}_{A(BC)}(\rho)$, 
which is what we want on the right-hand side.
On the left-hand
side we have two terms: (i) the average of the tangle between A and B
over a set of mixed states whose {\em average} is $\rho_{AB}$
({\em i.e.}, ${\rm Tr}_C (\rho)$), and (ii) the
average of the tangle between A and C over a set of 
mixed states whose average is
$\rho_{AC}$.  It is a fact that the tangle is a
{\em convex} function on the set of density matrices.\footnote{It
follows from Ref.~\cite{Wootters} that the concurrence is a
convex non-negative function on the set of density matrices for
two qubits.  The tangle, being the square of the
concurrence, is therefore also convex.}  That is, the average of
the tangles is greater than or equal to the tangle of the average.
In this case the tangles of the averages are $\tau_{AB} = 
\tau(\rho_{AB})$ 
and $\tau_{AC}=\tau(\rho_{AC})$.  The sum of these two tangles must
thus be less than or equal to $\tau^{min}_{A(BC)}(\rho)$, which is
what we wanted to prove.

Returning to the case of pure states, one may wonder how tight
the inequality (\ref{3taus}) is.  Could one find, for example, a
more stringent bound of the same form, based on a different measure
of entanglement?  To address this question, consider
the following pure state of ABC:
\begin{equation}
|\phi\rangle = \alpha |100\rangle + \beta |010\rangle + \gamma 
|001\rangle,
\label{phi}
\end{equation}
where the three positions in the kets refer to qubits A, B, and C
in that order.
For this state, one finds that $\tau_{AB}=4|\alpha|^2|\beta|^2$,
$\tau_{AC}=4|\alpha|^2|\gamma|^2$, 
and $\tau_{A(BC)}=4|\alpha|^2(|\beta|^2+|\gamma|^2)$.  Thus the 
inequality
(\ref{3taus}) becomes in this case an equality: 
$\tau_{AB}+\tau_{AC} = \tau_{A(BC)}$.  This example shows that for
any values of the tangles satisfying this equality, there is 
a quantum
state that is consistent with those values.  

Now let $T(\tau)$ be a monotonically
increasing function of $\tau$ that we might propose as an 
alternative
measure of entanglement.  For simplicity let us 
assume that $T(0)=0$ and $T(1)=1$.  
Because of the above example, $T$ could
satisfy the inequality 
$T_{AB} + T_{AC} \le T_{A(BC)}$ only if
$T(x) + T(y) \le T(x+y)$ for all non-negative $x$ and $y$
such that $x+y \le 1$.  Suppose $T$ has this property.
Then could there exist some quantum state for which
$T_{AB} + T_{AC} = T_{A(BC)}$ but $\tau_{AB}+\tau_{AC} < 
\tau_{A(BC)}$?
That is, could $T$ yield an equality for some state for which
$\tau$ gives only an inequality?  
The answer is no, because if $\tau_{AB}+\tau_{AC} < \tau_{A(BC)}$,
then $T_{AB} + T_{AC}$ = $T(\tau_{AB}) + T(\tau_{AC}) \le 
T(\tau_{AB}+\tau_{AC}) < T(\tau_{A(BC)}) = T_{A(BC)}$.  Moreover,
the only way $T$ can match $\tau$ in those cases where $\tau$
gives an equality is for $T$ to be equal to $\tau$.  In this sense,
$\tau$ is an optimal measure of entanglement with respect to
the inequality given in Eq.~(\ref{3taus}).  Note, however,
that the above argument applies only to functions of $\tau$.  
There could in principle be other measures of entanglement 
that are not
functions of $\tau$ that could make an equal claim to 
optimality.  

The entanglement of formation is a function of $\tau$, but
it is a concave function and therefore does not
satisfy an inequality of the form of Eq.~(\ref{3taus}).  
Consider, for example, the state $\frac{1}{\sqrt{2}}
|100\rangle + \frac{1}{2}|010\rangle + \frac{1}{2}|001\rangle$.
One finds that the relevant entanglements of formation are
$E_{AB} = 0.601$, $E_{AC} = 0.601$, and $E_{A(BC)} = 1$.
Thus, contrary to what one might expect, the 
sum of the entanglements
of formation between A and the separate qubits B and C is
greater than 1 ``ebit,'' despite the fact that 
one can reasonably regard 1 ebit as the
entanglement capacity of a single qubit.  This is not a paradox;
it simply shows us that entanglement of formation does not exhibit
this particular kind of additivity.  (This sense of ``additivity''
should not be confused with the additivity of entanglement when one
combines pairs to make larger systems\cite{Wootters}.
It is not known whether
entanglement of formation satisfies the 
latter notion of additivity.)

We have just seen that there are some states for which 
the inequality (\ref{3taus}) becomes an equality.  Of course
there are other states for which the inequality is strict.
As we will see, it turns out to be 
very interesting to consider the {\em difference}
between the two sides of Eq.~(\ref{3taus}).  This difference
can be thought of as the amount of entanglement between A and
BC that {\em cannot} be accounted for by the
entanglements of A with B and C separately.  In the following
paragraphs we refer to this quantity as the ``residual tangle.'' 

Let the system ABC be in a pure state $|\xi\rangle$, 
and as before, let
the components of $|\xi\rangle$ in the standard basis be $a_{ijk}$:
\begin{equation}
|\xi\rangle = \sum_{ijk}a_{ijk}|ijk\rangle.
\end{equation}
According to Eqs.~(\ref{2lambdas}) and (\ref{trdet}) 
and the discussion following 
Eq.~(\ref{first}), the residual tangle is equal to
\begin{equation}
\tau_{A(BC)} - \tau_{AB} - \tau_{AC} = 
2(\lambda^{AB}_1\lambda^{AB}_2 + \lambda^{AC}_1\lambda^{AC}_2),
\end{equation}
where $\lambda^{AB}_1$ and $\lambda^{AB}_2$ are the square roots of 
the two eigenvalues of $\rho_{AB}{\tilde{\rho}}_{AB}$, and 
$\lambda^{AC}_1$ and $\lambda^{AC}_2$ are defined similarly.
We now derive an explicit expression for the residual tangle 
in terms of the coefficients $a_{ijk}$.  

We focus first on the product
$\lambda^{AB}_1\lambda^{AB}_2$.  This product can almost be 
interpreted as the square root of the determinant of 
$\rho_{AB}{\tilde{\rho}}_{AB}$.  But 
$\rho_{AB}{\tilde{\rho}}_{AB}$ is an operator acting on a
four-dimensional space, and two of its eigenvalues are zero;
so its determinant is also zero.  However, if we consider
the action of $\rho_{AB}{\tilde{\rho}}_{AB}$ only on its 
{\em range}, then
$\lambda^{AB}_1\lambda^{AB}_2$ {\em will be} the square root of
the determinant of this restricted transformation. 

The range of $\rho_{AB}{\tilde{\rho}}_{AB}$ is spanned by
the two vectors $|v_0\rangle = \sum_{ij}a_{ij0}|ij0\rangle$ and
$|v_1\rangle = \sum_{ij}a_{ij1}|ij1\rangle$.  (These vectors
also span the range of $\rho_{AB}$.)  To examine the 
action of $\rho_{AB}{\tilde{\rho}}_{AB}$ on this subspace, we 
consider its effect on vectors of the form $x|v_1\rangle
+y|v_2\rangle \equiv \mtx{c}{x\\y}$.  This effect is given by
\begin{equation}
\mtx{c}{x^\prime \\ y^\prime} = R \mtx{c}{x\\y},
\end{equation}
where $R$ is a 2x2 matrix.  The product
$\lambda^{AB}_1\lambda^{AB}_2$
is the square root of the determinant of $R$.  
One finds that 
\begin{equation}
R_{ij} = \sum a_{klj}a^*_{mni}\sigma_{mp}
\sigma_{nq}a^*_{pqr}a_{str}\sigma_{sk}\sigma_{tl}, 
\end{equation}
where the sum is over all repeated indices.
(We have ordered the factors so as to suggest the expression
$\rho_{AB}{\tilde{\rho}}_{AB}$ from which $R$ is derived.)
Taking the determinant of $R$ involves somewhat tedious but
straightforward algebra, with the following result:
\begin{equation}
\lambda^{AB}_1\lambda^{AB}_2 = \sqrt{\det R} = |d_1 - 2d_2 + 4d_3|,
\end{equation}
where 
\begin{eqnarray}
d_1 & = & a^2_{000}a^2_{111}+a^2_{001}a^2_{110}+
      a^2_{010}a^2_{101}+a^2_{100}a^2_{011}; 
      \nonumber \\
d_2 & = & a_{000}a_{111}a_{011}a_{100} + 
    a_{000}a_{111}a_{101}a_{010} + 
    a_{000}a_{111}a_{110}a_{001} \\ 
    & + & a_{011}a_{100}a_{101}a_{010} +
    a_{011}a_{100}a_{110}a_{001} + 
    a_{101}a_{010}a_{110}a_{001}; \nonumber \\
d_3 & = & a_{000}a_{110}a_{101}a_{011} + 
    a_{111}a_{001}a_{010}a_{100}.  \nonumber 
\end{eqnarray}
We can get a mental picture of this expression by imagining the
eight coefficients $a_{ijk}$ attached to the corners of a cube.
Then each term appearing 
in $d_1$, $d_2$, or $d_3$ is a product of four of the coefficients 
$a_{ijk}$ such that the ``center of mass'' of the four
is at the center of the cube.  Such configurations fall into three 
classes:
those in which the four coefficients lie on a body diagonal and each
one is used twice
($d_1$); those in which
they lie on a diagonal plane ($d_2$), and those in which they lie
on the vertices of a tetrahedron ($d_3$).  Within each category,
all the possible configurations are given the same weight.   

This picture immediately yields an interesting fact: the quantity
$\lambda^{AB}_1\lambda^{AB}_2$ is invariant under permutations of the
qubits.  (A permutation of qubits corresponds to a reflection or
rotation of the cube, but each $d_i$ is invariant under such 
actions.)
This means in particular that we need not carry out a separate 
calculation
to find $\lambda^{AC}_1\lambda^{AC}_2$, since we know 
we will get the 
same result.  We can now therefore write down an expression for
the residual tangle:
\begin{equation}
\tau_{A(BC)} - \tau_{AB} - \tau_{AC} = 4|d_1 - 2d_2 + 4d_3|.
\label{residual}
\end{equation}
Note that the
residual tangle does not depend on which qubit one takes as the
``focus'' of the construction.  In our calculations we have focused
on entanglements with qubit A, but if we had chosen qubit B instead,
we would have found that
\begin{equation}
\tau_{B(CA)} - \tau_{BC} - \tau_{BA} = 4|d_1 - 2d_2 + 4d_3|.
\end{equation}
The residual tangle thus represents a collective property
of the three qubits that is unchanged by permutations; it is really
a kind of three-way tangle.  If we call this quantity $\tau_{ABC}$,
we can summarize the main results of this paper in the following
equation.
\begin{equation}
\tau_{A(BC)} = \tau_{AB} + \tau_{AC} + \tau_{ABC}. \label{final}
\end{equation}
In words, the tangle of A with BC is the sum of
its tangle with B, its tangle with C, and the essential
three-way tangle of the triple.  As an example, 
consider the Greenberger-Horne-Zeilinger state 
$\frac{1}{\sqrt{2}}(|000\rangle + |111\rangle)$ \cite{GHZ}.  
For this state the
tangle of each qubit with the rest of the system is 1,
the three-way tangle is also 1, and
all the pairwise tangles are zero (the qubits in each pair 
are classically correlated but not entangled).
Thus Eq.~(\ref{final}) in this case
becomes $1 = 0 + 0 + 1$.

Finally, we note that the expression for $\tau_{ABC}$ in terms of
$d_1$, $d_2$, and $d_3$ [Eq.~(\ref{residual})]
can be rewritten, after a little more 
algebra, in a more standard form:
\begin{equation}
\tau_{ABC} = 2\bigg|\sum a_{ijk}a_{i^\prime j^\prime m}
a_{npk^\prime}a_{n^\prime p^\prime m^\prime}
\epsilon_{ii^\prime}\epsilon_{jj^\prime}\epsilon_{kk^\prime}
\epsilon_{mm^\prime}\epsilon_{nn^\prime}\epsilon_{pp^\prime}
\bigg|,
\end{equation}
where the sum is over all the indices.  This
form does not immediately reveal the invariance of $\tau_{ABC}$
under permutations of the qubits, but the invariance
is there nonetheless.  

It would be very interesting to know which of the results 
of this paper generalize
to larger objects or to larger collections of objects.  
At this point it is not
clear how one might begin to generalize this approach to qutrits 
or higher
dimensional objects, because the spin-flip operation seems peculiar
to qubits.  On the other
hand, it appears very likely that at least some of these results 
can be
extended to larger collections of qubits.  The one solid piece of
evidence we can offer is the existence of a generalization of 
the state $|\phi\rangle$ of Eq.~(\ref{phi}) to $n$ qubits:
\begin{equation}
|\phi\rangle = \alpha_1|100\ldots0\rangle+\alpha_2|010\ldots0\rangle+
\alpha_3|001\ldots0\rangle+\cdots+\alpha_n|000...1\rangle.
\end{equation}
One can show that for this state, the following equality holds.
\begin{equation}
\tau_{12}+\tau_{13}+\cdots+\tau_{1n} = \tau_{1(23\ldots n)},
\end{equation}
where the qubits are now labeled by numbers rather than letters.
We are willing to conjecture that the corresponding inequality,
analogous to Eq.~(\ref{3taus}), is valid for all pure states
of $n$ qubits.

This work was supported in part by the National Science Foundation's
Research Experiences for Undergraduates program.  WKW is 
grateful for the support and hospitality of the
Isaac Newton Institute.  We would also like to thank
Tony Sudbery and Noah Linden for discussions that helped streamline the
derivation of Eq.~(\ref{first}).


\newpage

\begin{thebibliography}{99}

\bibitem{pure} C.~H.~Bennett, H.~J.~Bernstein, S.~Popescu, 
and B.~Schumacher, 
       {\em Phys.~Rev.~A} {\bf 53}, 2046 (1996).
\bibitem{entmeasures} C.~H.~Bennett, D.~P.~DiVincenzo, J.~Smolin, 
and W.~K.~Wootters, 
       {\em Phys.~Rev.~A} {\bf 54}, 3824 (1996).
\bibitem{relative} V.~Vedral, M.~B.~Plenio, M.~A.~Rippin, and 
P.~L.~Knight,
       {\em Phys. Rev. Lett.} {\bf 78}, 2275 (1997); V.~Vedral, 
       M.~B.~Plenio,
       K.~Jacobs, and P.~L.~Knight, {\em Phys. Rev. A} {\bf 56}, 
       4452 (1997);
       V.~Vedral and M.~B.~Plenio, {\em Phys. Rev. A} {\bf 57}, 
       1619 (1998).
\bibitem{Horodecki} M.~Horodecki, P.~Horodecki, and R.~Horodecki, 
{\em Phys. Rev. Lett.} {\bf 80}, 5239 (1998).
\bibitem{Rains} E.~M.~Rains, quant-ph/9809078, quant-ph/9809082.
\bibitem{Hill} S.~Hill and W.~K.~Wootters, {\em Phys. Rev. Lett.} 
{\bf 78}, 5022 (1997).
\bibitem{Wootters} W.~K.~Wootters, {\em Phys. Rev. Lett.} {\bf 80}, 
2245 (1998).
\bibitem{multi} The problem of quantifying multiparticle entanglement 
has been discussed, for example, in M.~Murao, M.~B.~Plenio, S.~Popescu, 
V.~Vedral, and P.~L.~Knight, {\em Phys. Rev. A} {\bf 57}, 4075 (1998)
and A.~Thapliyal, {\em Phys. Rev. A} {\bf 59}, 3336 (1999).
\bibitem{Linden} N.~Linden and S.~Popescu, quant-ph/9711017; 
N.~Linden, S.~Popescu, and A.~Sudbery, quant-ph/9801076.
\bibitem{Mahler} J.~Schleinz and G.~Mahler, {\em Phys. Lett. A} 
{\bf 224}, 39 (1996).
\bibitem{Beth} M.~Grassl, M.~R{\"o}tteler, and T.~Beth, 
quant-ph/9712040.
\bibitem{Kempe} J.~Kempe, quant-ph/9902036.
\bibitem{Buzek} V.~Bu\v{z}ek, V.~Vedral, M.~B.~Plenio, P.~L.~Knight,
and M.~H.~Hillery, {\em Phys. Rev. A} {\bf 55}, 3327 (1997).
\bibitem{Cerf} N.~Cerf, quant-ph/9805024.
\bibitem{cloning} D.~Bru\ss , D.~P.~DiVincenzo, A.~Ekert, C.~A.~Fuchs,
C.~Macchiavello, and J.~A.~Smolin, {\em Phys. Rev. A} 
{\bf 57}, 2368 (1998).
\bibitem{Karlsson} A.~Karlsson and M.~Bourennane, 
{\em Phys. Rev. A} {\bf 58}, 4394 (1998).
\bibitem{Murao} M.~Murao, D.~Jonathan, M.~B.~Plenio, and V.~Vedral,
{\em Phys. Rev. A} {\bf 59}, 156 (1999).
\bibitem{Bruss} D.~Bru\ss , quant-ph/9902023.
\bibitem{Royal} W.~K.~Wootters, {\em Phil. Trans. R. Soc. Lond. A} 
{\bf 356}, 1717 (1998). 
\bibitem{GHZ} D.~M.~Greenberger, M.~Horne, and A.~Zeilinger, in 
{\em Bell's 
Theorem, Quantum Theory, and Conceptions of the Universe}, ed. 
M.~Kafatos
(Kluwer 1989).

\end{thebibliography}
\end{document}